\documentclass[12pt]{article}
\usepackage{float}
\usepackage{natbib}
\usepackage{multirow}
\usepackage{multicol}
 \usepackage{graphicx}
 \usepackage{mathtools}
 \usepackage{listings}
 \usepackage{color}
 \usepackage{setspace}
 \usepackage{pdfpages}
 \usepackage{amsmath}
 \usepackage{amsthm}
 \usepackage{amssymb}
 \usepackage{subfig}
 \usepackage{tabularx}
\usepackage{bm}
  \usepackage{url}
  \usepackage{authblk}
  
\newcommand{\blind}{0}
\newcommand\indpt{\protect\mathpalette{\protect\independenT}{\perp}}
\def\independenT#1#2{\mathrel{\rlap{$#1#2$}\mkern2mu{#1#2}}}

\newcommand{\Yb}{\bm{Y}}
\newcommand{\Xb}{\bm{X}}
\newcommand{\Ab}{\bm{A}}
\newcommand{\Bb}{\bm{B}}
\newcommand{\mub}{\bm{\mu}}

\newcommand{\diag}{\mbox{diag}}
\newcommand{\GG}{\mathcal{G}}
\newcommand{\ul}[1]{\underline{#1}}

\usepackage{xcolor}         %for colors

\def\hei{\color{black}}

\newcommand{\ech}{\hei \rm}

\begin{document}
	\def\spacingset#1{\renewcommand{\baselinestretch}%
		{#1}\small\normalsize} \spacingset{1}

	%%%%%%%%%%%%%%%%%%%%%%%%%%%%%%%%%%%%%%%%%%%%%%%%%%%%%%%%%%%%%%%%%%%%%%%%%%%%%%

	\if0\blind
	{
		\title{\bf Reciprocal Graphical Models for Integrative Gene Regulatory Network Analysis}
		\author[1]{Yang Ni}
		\author[2]{Yuan Ji}
		\author[3]{Peter M\"uller}
		\affil[1]{Department of Statistics and Data Sciences, The University of Texas at Austin}
		\affil[2]{Program for Computational Genomics \& Medicine, NorthShore University HealthSystem\\
			Department of Public Health Sciences, The University of Chicago}
		\affil[3]{Department of Mathematics, The University of Texas at Austin}
%		\author{Yang Ni\hspace{.2cm}\\
%			Department of Statistics and Data Sciences, The University of Texas at Austin\\
%			and \\
%			Yuan Ji \\
%			Program for Computational Genomics \& Medicine, NorthShore University HealthSystem\\
%			Department of Public Health Sciences, The University of Chicago\\
%			and \\
%			Peter M\"uller \\
%			Department of Mathematics, The University of Texas at Austin}
		\date{}
		\maketitle
	} \fi
	
	\if1\blind
	{
		\bigskip
		\bigskip
		\bigskip
		\begin{center}
			{\LARGE\bf Reciprocal Graphical Models for Integrative Gene Regulatory Network Analysis}
		\end{center}
		\medskip
	} \fi
	
\bigskip
\begin{abstract} Constructing gene regulatory networks is a
  fundamental task in systems biology. We introduce a Gaussian
  reciprocal graphical model for inference about gene regulatory
  relationships by integrating mRNA gene expression and DNA level
  information including copy number and methylation. Data integration
  allows for inference on the directionality of certain regulatory
  relationships, which would be otherwise indistinguishable due to
  Markov equivalence. Efficient inference is developed based on
  simultaneous equation models. Bayesian model selection
  techniques are adopted to estimate the graph structure. We
  illustrate our approach by simulations and two applications in ZODIAC
  pairwise gene interaction analysis and colon adenocarcinoma pathway
  analysis.
\end{abstract}
\noindent%
\vfill
\newpage
\spacingset{1.45} 
% DON'T change the spacing!

\section{Introduction}
In this paper, we develop a reciprocal graphical model (RGM) to infer
gene regulatory relationships and gene networks.
This includes in particular directed edges without time course or
interventional data.  RGMs allow for 
undirected edges, directed edges and directed cycles and therefore are
ideally suited for modeling regulatory relationships including feedback
loops.
Exploiting genomic
data from multiple modalities/platforms, we are able to determine the
directionality of certain regulatory relationships, which would be
otherwise indistinguishable due to Markov equivalence. 
Such inference about directionality becomes possible because 
basic biology fixes the directionality for some edges, for example,
between   DNA   methylation and gene expression   of the same
gene.   Conditioning on such known
directionality enables us to infer directionality for other 
edges. 
Statistically, the class of probability models determined by RGMs is
strictly larger than the class of probability models determined by
directed acyclic graphs (DAGs) and Markov random fields
(MRFs). Computationally, the connection of RGMs with
simultaneous equation models (SEMs) facilitates computation-efficient
implementation of full posterior inference.

% Gene regulatory networks (GRNs) describe the regulatory relationships
% between genes and their products. Regulators govern the gene
% expression of mRNAs and proteins, which plays an important role in
%  many biologic processes, 
% \note{``aspects of life'' sounds too poetic :-)} including cell
% differentiation, metabolism, the cell cycle and signal transduction
% \citep{karlebach2008modelling}.
% \cut
% Constructing GRNs is therefore a fundamental task in
% systems biology. However, it is a challenging inverse problem because
% the network structure is latent and only gene expressions are
% observed.

Most recent graphical model approaches in biostatistics and
bioinformatics are restricted to DAGs  
\citep{stingo2010bayesian,yajima2015detecting,ni2015bayesian} and MRFs
\citep{wang2009bayesian,dobra2012bayesian,green2013sampling,mitra2013bayesian,
     Wang2013}. 
These approaches use the conditional independence
structure represented by the graphical models. 
% to capture the regulatory mechanism in GRNs probabilistically, 
The  common use of  MRFs and DAGs is due to 
mathematical  tractability and easy computation, despite some
inherent limitations. 
MRFs  model  undirected  relationships between genes and do
not account for the directionality of edges.
 However,  biological interactions between genes are often
asymmetric.
For example, it is only possible for a regulator to regulate its targets,
not vice versa. DAGs allow for directed edges but
 only as arbitrary factorization of a joint probability model. 
Also, DAGs  explicitly prohibit
directed cycles. 
 However,  feedback loops are quite common motifs and have key
functional roles in many cellular processes such as regulating gene
expressions and acting as bistable switches
\citep{shin2010functional}.

% In this paper, we  use  a reciprocal graphical model (RGM)
% to infer gene regulatory relationships. RGMs allow for undirected
% edges, directed edges and directed cycles. 
% \note{avoid long sentences -- if it's easy just split them...}
% RGMs are therefore suitable
% for modeling various regulatory relationships including feedback
% loops. Statistically, the class of probability models determined by
% RGMs is strictly larger than the class of probability models
% determined by DAGs and MRFs. We exploit the connection of RGMs with
% simultaneous equation models (SEMs) which renders efficient inference
% algorithm.

RGMs were first proposed by \cite{koster1996markov} but 
 remain curiously under-used   in the biostatistics and
bioinformatics literature, with few exceptions.
\cite{zhang2005hierarchical} developed a hierarchical RGM
to discover the relationship between cholesterol levels and pulmonary
function in a longitudinal study. The graph structure is fixed and a
generalized EM algorithm was developed to estimate the parameters.
\nocite{telesca2012modeling2,telesca2012modeling1}
 Telesca {\it et al.} (2012ab) 
use RGMs for latent variables that represent active/inactive
proteins  and differential vs. non-differential gene
expression, respectively.  However, the use of the RGM is restricted
to representation and for convenient summaries. The actual inference
model is based on the implied conditional independence structure only
(after moralization). Also, they use RGMs with only directed edges,
excluding possible undirected edges. 
% \cite{telesca2012modeling1} built a hierarchical model to
% study dependence structure of gene expressions. They first model the
% observed gene expressions by a three-component mixture model and
% introduce trinary latent variables indicating low, normal and high
% expressions. The latent variables are then specified as a
% deterministic function of latent Gaussian variables for which the
% prior model is given by RGM. Although the graph structure is not fixed
% but learned from the data, their model is an indirect approach for
% modeling dependence of gene expressions in that it is unclear how the
% dependence structure of latent Gaussian variables is passed through
% the hierarchies to the level of observed gene expressions. 
By contrast, our use of RGMs is directly on the observed gene
expressions and  we use the RGM to build the probability model in
a way that allows us to infer direction for directed edges.

Because different RGMs may determine  equivalent 
% statistical models, that is, models with the same
conditional independence structure, inference algorithms that
use only the implied conditional independence structure can not
possibly distinguish such equivalent graphs based on observational
data. In particular,  
% ignore this identifiability issue are unable to differentiate these
% RGMs from an observational study, for example,
the directionality of some
regulatory relationships cannot be determined (examples are given in
Section \ref{sec:rgm}).  \ech
% None of the aforementioned approaches
% addresses this fundamental problem explicitly. 
In this paper, we propose  a possible solution. 
We integrate different sources of genomic information including DNA
copy number and DNA methylation in a way that allows us to determine
the direction of some edges from first biological principles. This
allows us then to
identify the
gene regulatory networks (GRN) that best fit the data, including
inference on edge directions in some cases. 
That is,  the additional genetic and epigenetic information
together with fundamental biological knowledge can inform the
directionality of the regulatory relationships between genes.

\indent 
The proposed approach is motivated by two interesting applications. In
the first application, we extend ZODIAC
\citep{zhu2015zodiac} to incorporating directed edges between
genes. ZODIAC is a free online tool
(\url{http://www.compgenome.org/ZODIAC}) that provides (undirected)
molecular interactions for about 200 million pairs of genes across
different cancers based on multimodal genomic data from The Cancer
Genome Atlas (TCGA). It can be used to explore gene regulations in
cancer. With our contribution, ZODIAC will enable researchers to not
only find out whether certain genetic interactions are likely to be
present in cancer but also identify the sources (e.g. transcription
factors) and targets (e.g. targeted genes) of the regulations. Our
second application studies GRNs of colon adenocarcinoma (COAD). We
focus on genes from the RAS-MAPK pathway which is critical in the
initiation and progression of COAD
\citep{cancer2012comprehensive}. Integrating copy number and
methylation information, we find biologically meaningful gene
interactions as well as novel regulatory relationships that need to be
validated by further biological experiment. 

Finally, for clarification and to avoid misunderstanding
we note that our use of
graphical models is quite different from inference for random graph models
\citep{hoff_2009_cmot,goldenberg2010survey} when 
the graph itself is the data.

\indent The rest of the article is organized as follows. We describe our model in Section \ref{sec:model}. We present simulation studies in Section \ref{sec:sim} and two applications in Section \ref{sec:app}. Section \ref{sec:disc} provides our closing discussion.

\section{Model}
\label{sec:model}
\subsection{ Notation}
A graph $\GG=(V,E)$ consists of a set of \textit{vertices}
$V=\{1,\dots,p\}$ and a set of \textit{edges} $E$ connecting these
vertices. We consider both \textit{directed} and \textit{undirected}
edges $E=E^d\cup E^u$ with $E^d\subseteq\{(i,j)\mid i,j\in V\}$ and
$E^u\subseteq\{\{i,j\}\mid i,j\in V\}$ where the ordered pair $(i,j)$
denotes a directed edge from vertex $i$ to vertex $j$ and
$\{i,j\}$ denotes an undirected edge. We use the vertices to index a set of random variables, 
$\Yb=\Yb_V=(Y_1,\dots,Y_p)^T$.
For example, in our application $Y_j$ represents gene expressions for
gene $j$.

\begin{figure}[h]
	\centering
		\begin{tabular}{cccc}
	\subfloat[]{\includegraphics[width=0.2\textwidth]{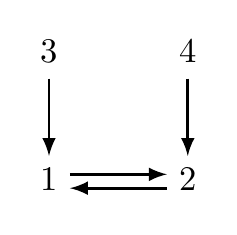}\label{rgm_fig}}& 
	\subfloat[]{\includegraphics[width=0.2\textwidth]{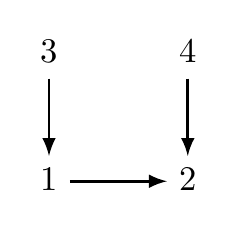}\label{dag1}}&
	\subfloat[]{\includegraphics[width=0.2\textwidth]{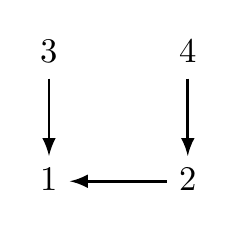}\label{dag2}} &
	\subfloat[]{\includegraphics[width=0.2\textwidth]{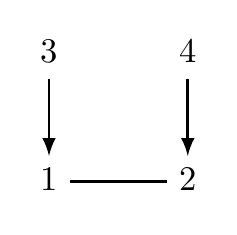}\label{ug_fig}}\\
	\subfloat[]{\includegraphics[width=0.2\textwidth]{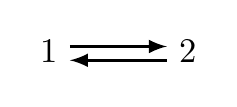}\label{tg1}} &
	\subfloat[]{\includegraphics[width=0.2\textwidth]{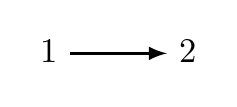}\label{tg2}}&
	\subfloat[]{\includegraphics[width=0.2\textwidth]{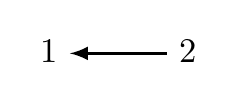}\label{tg3}} &
	\subfloat[]{\includegraphics[width=0.2\textwidth]{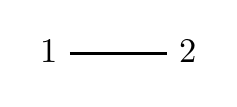}\label{tg4}}
	\end{tabular}
	\caption{Graphs for illustration.   Graphs in (a) -- (d) are
          reciprocal graphs and 
          imply different conditional independence. Graphs in (e) --
          (h) are Markov equivalent as they have imply the same Markov
        properties. In our applications, we will use nodes 1 and 2 to
        represent gene expression of two genes and nodes 3 and 4 to
        represent copy number (or methylation) of the same genes.   }\label{rdu}
\end{figure}

A \textit{path}
of length $K$ is an ordered sequence $(i_0,\dots,i_K)$ of distinct
vertices except possibly $i_0=i_K$ such that $\{i_{k-1},i_k\}\in E$ or
$(i_{k-1},i_k)\in E$ for $k=1,\dots,K$.  A path  is called undirected if
$\{i_{k-1},i_k\}\in E$ for  all  $k=1,\dots,K$.  A \textit{path component}
is a set of vertices that are all connected by an undirected
path. A \textit{reciprocal graph} (RG) is a graph
$\GG=(V,E)$ such that there are no directed edges between
vertices in the same path component \citep{koster1996markov}.
Some examples  of RG with four
vertices are given in Figure \ref{rgm_fig}. The \textit{boundary} of a
vertex $i$ is $\mbox{bd}(i)=\{j\mid\{j,i\}\in E\mbox{~or~}(j,i)\in E\}$
and the \textit{boundary} of a subset $V_0\subseteq V$ is
$\mbox{bd}(V_0)=\bigcup_{i\in V_0}\mbox{bd}(i)\backslash V_0$. An
anterior set is a subset $V_0\subseteq V$ such that
$\mbox{bd}(V_0)=\emptyset$ and the smallest anterior set containing
$V_0$ is denoted by $\mbox{an}(V_0)$. 

The Markov property
(i.e. conditional independence relationships of $\Yb$) of an RG relies on the
notion of \textit{moralization}. To moralize a graph $\GG$, we
connect all vertices in the boundary of each path component of
$\GG$ by undirected edges and then replace all directed edges
by undirected edges. The resulting moral graph is an undirected graph
and is denoted by $\GG^m$. 
Later we will use graph \textit{separation} 
to introduce a global Markov property. In an undirected graph, two
sets $V_1$ and $V_2$ are said to be separated by a third set $V_3$ if
every path between $V_1$ and $V_2$ intersects $V_3$. 
\subsection{Reciprocal graphical models, Markov equivalence and integrative genomics}
\label{sec:rgm}
\indent A graphical model is a mapping between a family of
distribution and an underlying graph. 
% (i.e., its encoded Markov properties). 
In this paper, we focus on the class of RGs which by
definition strictly contains MRFs and DAGs as subclasses.

The probability distribution of
$\Yb$ is said to be (global) \textit {Markov} with respect to an RG
$\GG$ if $\Yb_{V_1}\indpt \Yb_{V_2}\mid\Yb_{V_3}$
whenever $V_3$ separates $V_1$ and $V_2$ in $\GG_{an(V_1\cup
  V_2 \cup V_3)}^m$. 
 RGMs can represent global Markov properties beyond the
conditional independence structure that is encoded in MRFs and DAGs. 
RGMs are a strictly larger
class of probability model than MRFs and DAGs. For example, in Figure
\ref{rgm_fig}, the only two conditional independence relationships are
$3\indpt 4$ and $3\indpt4\mid1,2$. There is no MRF or DAG that encodes
the same conditional independence relationships.
More importantly for our application, 
RGMs are % extends the interpretation of MRFs and DAGs and is
particularly useful for the construction of genomic networks because
of the
ability in modeling feedback loops which are not allowed by MRFs or
DAGs. For example, in Figure \ref{rgm_fig}, gene 1 may regulate the
expression of gene 2 while the status of gene 2 may reciprocally
affect gene 1 through a feedback regulatory mechanism.

Two graphs are said to be \textit{Markov equivalent} if they
have the same Markov properties. For example, graphs \ref{tg1},
\ref{tg2}, \ref{tg3} and \ref{tg4} are Markov equivalent as they have
the same Markov property, namely, no conditional independence
assertion. Markov equivalence is indeed an equivalence relation and
hence induces Markov equivalence class. Graphs within the same Markov
equivalence class are usually nonidentifiable from observational
data. In particular, observational data do not allow for inference on the direction of the edges in Figures \ref{tg1}-\ref{tg4}. 

However, with prior knowledge, sometimes we are able to
distinguish the relationships between variables. In this paper,
we introduce such prior knowledge by
deliberately considering edges with (biologically) known
direction if included. 
This allows for inference on other edges. 
We develop this approach based on
integrating genomic information across different platforms and exploiting
the central dogma of molecular biology that mRNA is produced by
transcription from segments of DNA on which the copy number and
methylation are measured, but the reverse processes are rare and
biologically uninterpretable. For illustration, let vertices 1 and 2
in Figure \ref{rdu} represent mRNA gene expressions and vertices 3 and
4 represent copy numbers of the corresponding genes and assume there
exists dependence between copy number and gene expression for each
gene and hence by the central dogma, there are directed edges from 3
to 1 ($3\rightarrow1$) and from 4 to 2 ($4\rightarrow 2$). Now we are
able to fully identify the relationship between mRNAs 1 and 2 because
each of graphs \ref{rgm_fig}, \ref{dag1}, \ref{dag2} and \ref{ug_fig}
defines a distinct set of conditional independence relationships.

%\begin{figure}
%	\centering
%	\subfigure[]{\includegraphics[width=0.2\textwidth]{g3.pdf}\label{tg1}} 
%	\subfigure[]{\includegraphics[width=0.2\textwidth]{g1.pdf}}
%	\subfigure[]{\includegraphics[width=0.2\textwidth]{g2.pdf}} 
%	\subfigure[]{\includegraphics[width=0.2\textwidth]{g4.pdf}}
%	\caption{}\label{twogenes}
%\end{figure}
\subsection{Simultaneous equation models}
\begin{figure}[h]
	\centering
	\includegraphics[width=.5\textwidth]{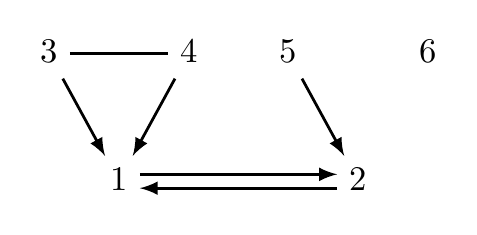}
	\caption{Path diagram for simultaneous equations.}\label{fig:sem}
\end{figure}

For a formal description of our approach, we still need the mapping
from the RGM to a family of probability models for the observational
data. Let $\Yb=(Y_1,\dots,Y_p)^T$
denote the mRNA gene expressions for genes $1,\dots,p$. Let
$\Xb=(X_1,\dots,X_{2p})^T$ be the set of DNA level measurements for
genes $1,\dots,p$ with $X_{2i-1}$ and $X_{2i}$ being the copy number
and the methylation for gene $i$, respectively. We first state
the simultaneous equation model (SEM) and will then introduce the
mapping. An SEM for $\Yb$ and $\Xb$ is given by
\begin{eqnarray}
  \label{eqn:sem}
  \Yb=\bm{A}\Yb+\bm{B}\bm{X}+\bm{E}
\end{eqnarray}
where $\bm{A}=(a_{ij})\in \mathbb{R}^{p\times p}$ with zeros on the
diagonal, $\bm{B}=(b_{ij})\in \mathbb{R}^{p\times 2p}$,
$\bm{E}=(\epsilon_1,\dots,\epsilon_p)^T\sim
\mbox{N}_p(0,\bm{\Sigma})$, $\bm{X}\sim
\mbox{N}_{2p}(\bm{0},\bm{\Psi})$ and $\bm{E}$ and $\bm{X}$ are
independent. Provided $\bm{I}_p-\bm{A}$ is invertible where $\bm{I}_p$
is a $p\times p$ identity matrix, model (\ref{eqn:sem}) can be
equivalently expressed as 
\begin{eqnarray}
\label{ygx}
\Yb\mid\bm{X}&\sim& \mbox{N}_p\left\{(\bm{I}_p-\bm{A})^{-1}\bm{B}\bm{X},(\bm{I}_p-\bm{A})^{-1}\bm{\Sigma}(\bm{I}_p-\bm{A})^{-T}\right\},\\
\label{xmar}
\bm{X}&\sim& \mbox{N}_{2p}(\bm{0},\bm{\Psi}).
\end{eqnarray}
To link an SEM to an RGM, we draw a path diagram $\GG=(V,E)$ of SEM by the following rules:
\begin{itemize}
	\item[(i)] define vertices $V=\{1,\dots,p,p+1,\dots,3p\}$ which represent $(\Yb,\bm{X})=(Y_1,\dots,Y_p,X_1,\dots,X_{2p})$;
	\item[(ii)] draw directed edges $E^d=\{(j,i)\mid a_{ij}\neq 0 \mbox{~or~} b_{i,j-p}\neq 0\}$; and
	\item[(iii)] draw undirected edges $E^u\{\{i,j\}\mid \bm{\Psi}_{i-p,j-p}\neq 0 \}$.
\end{itemize}
In words, (i) we introduce a node for each variable in $(\bm{Y},\bm{X})$ with nodes $j=1,\dots,p$ corresponding to $Y_j$ and $p+j$ corresponding to $X_j$ for $j=1,\dots,2p$; (ii) nodes $i=1,\dots,p$ (i.e. $Y_i$ nodes) become targets of directed edges from node $j$ if the corresponding $a_{ij}\neq 0$ or $b_{i,j-p}\neq 0$; (iii) we introduce undirected edges between $X_i$ and $X_j$ (i.e. nodes $i+p$ and $j+p$) if $\bm{\Psi}_{i,j}\neq 0$.
 Let $A=\diag(A_1,A_2)$ denote a block diagonal matrix with
diagonal blocks $A_1$ and $A_2$. 
Figure \ref{fig:sem} shows an example  of an RGM  with $p=2$, 
$$
  \bm{A}=\left[\begin{array}{cc}0&*\\ *&0\end{array}\right],~
  \bm{B}=\left[\begin{array}{cccc}*&*&0&0\\0&0&*&0\end{array}\right]
\mbox{ and }
\bm{\Psi}=\diag
\left(\left[\begin{array}{cc}*&*\\
              *&*\end{array}\right],\bm{I}_2\right),
$$
with $*$ indicating non-zero elements. In general,  the path
  diagram of an SEM following rules (i)-(iii) is not always an RGM.
However, the probability 
distribution of $(\Yb,\bm{X})$ defined by a SEM is Markov with respect
to the path diagram $\GG$
\citep{spirtes1995directed,koster1996markov} if $\bm{\Sigma}$ is diagonal
$\bm{\Sigma}=\mbox{diag}(\sigma_1,\dots,\sigma_p)$ and $\bm{\Psi}$ is
block diagonal with each block being a full matrix (without zeros). Since the main
inferential goal of this paper is to investigate the regulatory
relationship between genes, we will not model the marginal distribution
of $\Xb$ and will only focus on the conditional distribution of
$\Yb\mid \Xb$. We fix $b_{ij}$ to 0 for $j\neq 2i-1$ or $2i$ because
copy number and methylation of gene $i$, in principle, only affect the
expression of gene $i$. 

\subsection{Priors and graph structure learning}
The structural zeros in $\bm{A}$ and $\bm{B}$ correspond to missing
edges in the RG. Therefore, learning the graph structure is equivalent
to finding sparse estimators for $\bm{A}$ and $\bm{B}$. Towards this
end, we define a non-local prior for $\bm{A}$ and $\bm{B}$,
which is constructed as follows. We put a thresholded prior on each
element of $\bm{A}$ and $\bm{B}$. We first write $a_{ij}$ and
$b_{ik}$ as 
$$
  a_{ij}=\tilde{a}_{ij}\mbox{I}(\mid\tilde{a}_{ij}\mid>t_i)\mbox{~~and~~}b_{ik}=\tilde{b}_{ik}\mbox{I}(\mid\tilde{b}_{ik}\mid>t_i),
$$
for $i=1,\dots,p$, $j\neq i$ and $k=2i-1,2i$. The threshold parameter
$t_i$ controls the minimum effect sizes of $a_{ij}$ and $b_{ik}$. We
do not fix $t_i$ but instead learn it from the data by assigning a
prior $t_i\sim \mbox{Uniform}(0,t_0)$. We impose normal priors for
$\tilde{a}_{ij}$ and 
$\tilde{b}_{ik}$, $\tilde{a}_{ij}\sim \mbox{N}(0,\tau_{ij})$ and 
$\tilde{b}_{ik}\sim \mbox{N}(0,\nu_{ik})$ and conjugate hyperpriors 
$\tau_{ij}\sim \mbox{IG}(\alpha_\tau,\beta_\tau)$ and $\nu_{ik}\sim
\mbox{IG}(\alpha_\nu,\beta_\nu)$. 
\cite{ni2016} show that marginally, after integrating with respect to
$t_i$, the induced
marginal prior for $a_{ij}$ or $b_{ik}$ is a mixture of a
point mass at 0 and a non-local prior \citep{johnson2010use}. 
Non-local priors shrink small effects
to zero, which is desirable in our setting where we are only interested
in edges with moderate to strong effects. We complete our prior
specifications with a conjugate prior $\sigma_i\sim
\mbox{IG}(\alpha_\sigma,\beta_\sigma)$.

Posterior inference is straightforward by Markov chain Monte
Carlo (MCMC). It is worth mentioning that we do not need to check the
invertibility of $\bm{I-A}$ because $\bm{I-A}$ is practically always
invertible and we never have to invert the matrix in our MCMC. Details
are provided in Supplementary Material A. Let
$\bm{\Gamma}_A=(\gamma_{A,ij})$ and 
$\bm{\Gamma}_B=(\gamma_{B,ik})$ denote matrices indicating non-zero
elements in $\bm{A}$ and $\bm{B}$, respectively. 
An estimated graph can be reported by 
selecting edges for which the marginal
posterior probability $p(\gamma_{A,ij}=1\mid\mbox{Data})$ or
$p(\gamma_{B,ik}=1\mid\mbox{Data})$ exceeds a certain cutoff. 
This is the median probability model (MPM) when the cutoff is fixed at
1/2. Instead of being fixed at an arbitrary value, the cutoff can be
chosen to control the posterior expected false discovery rate (FDR) 
at a desired level $\alpha$
\citep{newton2004detecting,muller2006fdr}. Alternatively we report the highest posterior probability model
(HPM)
by maximizing 
$p(\bm{\Gamma}_A,\bm{\Gamma}_B\mid\mbox{Data})$. However, this is only
feasible when the model space is small (i.e. with very small
$p$). 

\section{Simulations}
\label{sec:sim}
We carry out a simulation study to validate the model's ability
to recover a true graph with sample sizes similar to the later
application. We generate data
by mimicking the actual data from the later application, using
a sample size of $n=276$ and   $p=10$  genes. For each gene, we
generate two hypothetical DNA level measurements $\Xb$ from
\eqref{xmar} with $\bm{\Psi}=\bm{I}_{2p}$ (corresponding to copy
number and methylation in the real data). 

We consider three scenarios. In \ul{scenario 1}, we randomly set
$4/5$ entires in $\bm{A}$ and $1/3$ entries in $\bm{B}$ to zero. To
ensure that the true model is identifiable, we 
restrict the simulation truth to include 
at least one edge from DNA level measurements to each gene
(this is relaxed in scenario 2). We draw nonzero elements of $\bm{A}$ and
$\bm{B}$ from $\{-0.5,0.5\}$ with equal probability and let 
$\bm{\Sigma}=0.25\bm{I}_{10}$ in the simulation truth.

In \ul{scenario 2}, we reduce the signal by drawing nonzero
elements of $\bm{A}$ and $\bm{B}$ from $\{-0.4,0.4\}$ and increase the
noise  to $\bm{\Sigma}=\bm{I}_{10}$. Moreover, we do not force every
gene to be connected to at least one of its DNA level measurements.
As a result one gene is independent of both its DNA level
measurements. For scenarios 1 and 2, $\Yb$ is then generated from (\ref{ygx}).  

\ul{Scenario 3} is the same as scenario 1, except that the conditional
distribution of $\Yb\mid\bm{X}$ is misspecified and generated from a
$p$-dimensional multivariate t-distribution
$\mbox{T}_p(\mub,\bm{\Theta},\delta)$ with location
$\mub=(\bm{I}_p-\bm{A})^{-1}\bm{B}\bm{X}$, scale matrix
$\bm{\Theta}=(\bm{I}_p-\bm{A})^{-1}\bm{\Sigma}(\bm{I}_p-\bm{A})^{-T}$
and degrees of freedom $\delta=3$. 
By design, scenarios 2 and 3 are more difficult and perhaps more
realistic than scenario 1.

The hyperparameters are specified as
$\alpha_\sigma=\beta_\sigma=\alpha_\tau=\beta_\tau=\alpha_\nu=\beta_\nu=0.01$
and $t_0=1$. We run 50,000 MCMC iterations, discard the first 25,000
iterations as burn-in and retain every 5th sample. The graph is
estimated as an MPM.

\begin{figure}[h]
	\centering
	\subfloat[Scenario 2]{\includegraphics[width=0.4\textwidth]{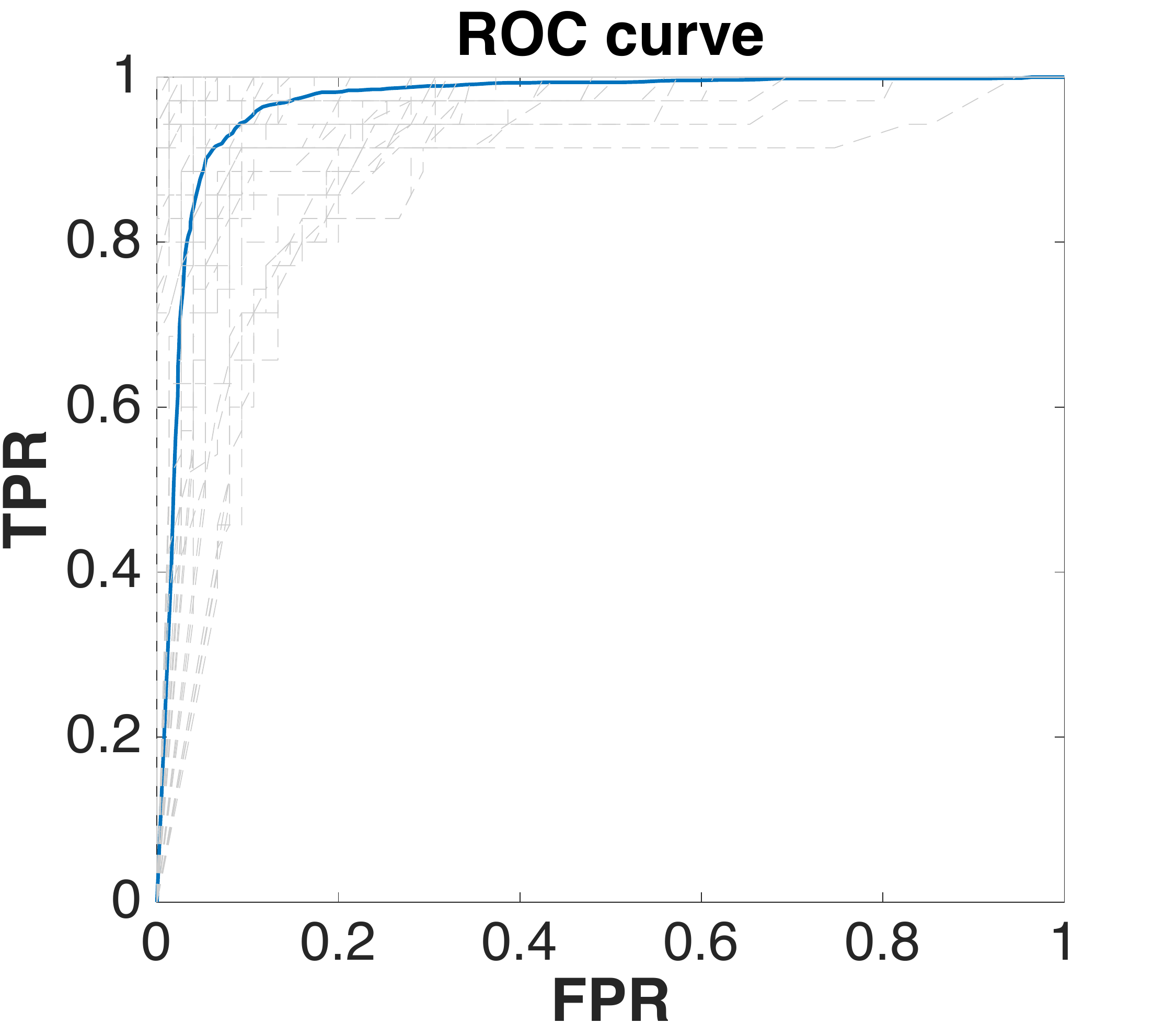}} 
	\subfloat[Scenario 3]{\includegraphics[width=0.4\textwidth]{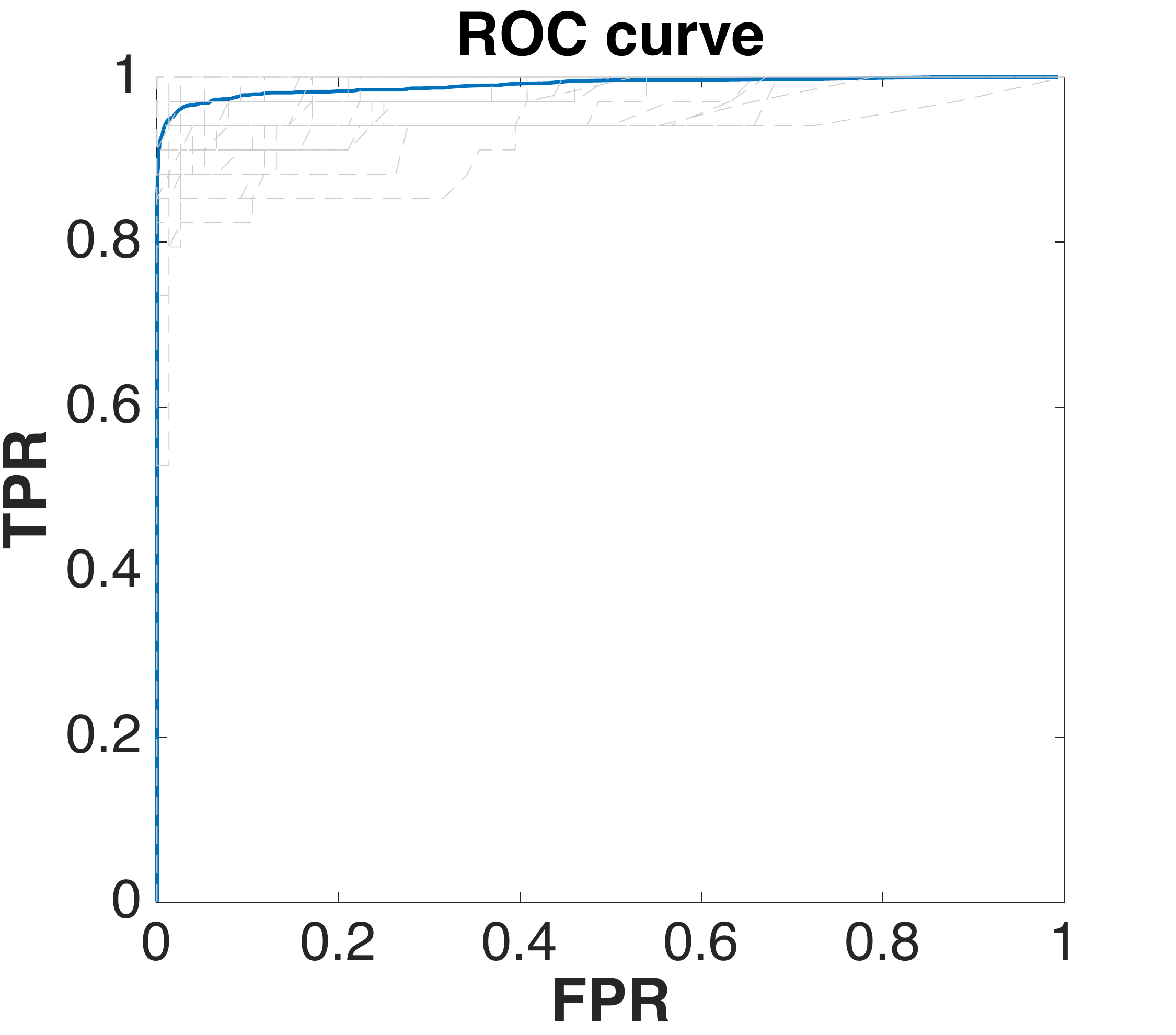}}
		\caption{ROC curves for scenarios 2 (panel a) and 3 (panel b). Each dashed line is an ROC curve for one simulation and the solid line is the average ROC curve over 50 simulations.}\label{simroc}
\end{figure}

In Table \ref{simres} we report the true positive rate (TPR), FDR and Matthews correlation
coefficient (MCC). 
We also plot the receiving
operating characteristic (ROC) curves in Figure \ref{simroc}. The ROC curve, TPR and FDR are for classifying 
entries in $\Ab$ and $\Bb$ as non-zero (``positive'') or not
(``negative''). That is, for the inclusion of all possible edges. 
TPR and FDR are the realized error rates under the reported MPM graph. The area under the ROC curve (AUC) is reported in Table \ref{simres} (the
ROC plot for scenario 1 is not shown -- the AUC is 1). The
performance in scenario 1 is nearly perfect. As expected, 
scenarios 2 and 3 are more challenging. In particular, 
we find a higher FDR, due to the lower signal-to-noise ratio and a minor
identifiability issue for scenario 2 and model misspecification for
scenario 3.

\begin{table}[h]
\caption{Simulation results. The average operating characteristics are calculated on the basis of 50 simulations; standard deviations are given within parentheses.}
\begin{center}
\begin{tabular}{ccccc}
\hline\hline
Scenario&MCC&TPR&FDR&AUC\\\hline
1&0.99 (0.02) & 1.00 (0.00)  &  0.01 (0.02) &  1.00 (0.00)\\
2&0.76 (0.13)&   0.98 (0.03) &   0.26 (0.14)  &  0.97 (0.04)\\
3&0.87 (0.08)&   0.97 (0.04) &   0.14 (0.08)  &  0.99 (0.02)\\\hline
\end{tabular}
\end{center}
\label{simres}
\end{table}%
\section{Application}
\label{sec:app}
\subsection{ZODIAC pairwise gene interaction analysis}
\label{sec:zodiac}
We use TCGA-Assembler \citep{zhu2014tcga} to retrieve mRNA gene
expression  (GE), DNA copy number (CN) and DNA methylation (ME) data
for different cancer types from TCGA. For illustration, we report the
analysis based on pairs of genes whose GEs are moderately correlated
(between 0.5 and 0.7) and show high correlations ($>0.8$) with
corresponding CNs or MEs. The resulting dataset consists of 1395 pairs
from 15 cancers (details of the data are in secton B of the Supplementary
Material). We apply our method separately to each pair of genes using
the same hyperparameter and MCMC settings as in the simulation study.
Since $p=2$ is sufficiently small, we can report the HPM as an
estimate of the graph.  
In this application, we aim to demonstrate: (1) like
ZODIAC, our model can be used as a hypothesis generating tool; 
(2) compared to the undirected graph model used in ZODIAC, the
RGM is more flexible and allows for more and different inference and
interpretation. 
% unlike ZODIAC, our model is more flexible and is interpreted
% differently. \\

\begin{figure}[h]
	\centering
	\subfloat[READ]{\includegraphics[width=0.2\textwidth]{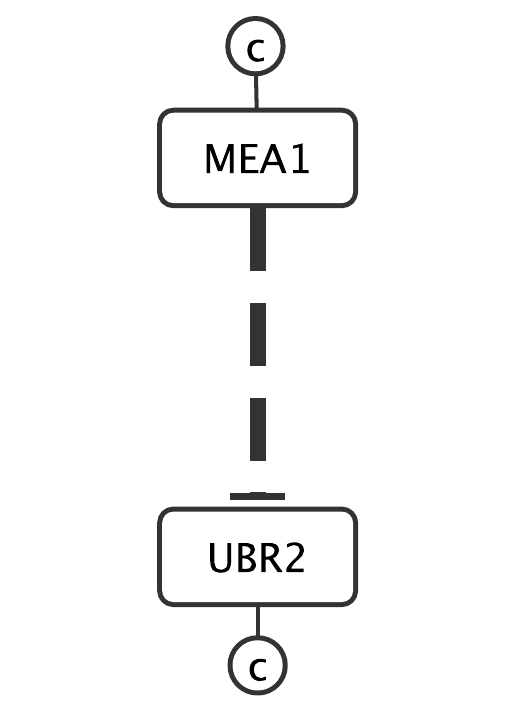}\label{zod2}}	\subfloat[TGCT]{\includegraphics[width=0.2\textwidth]{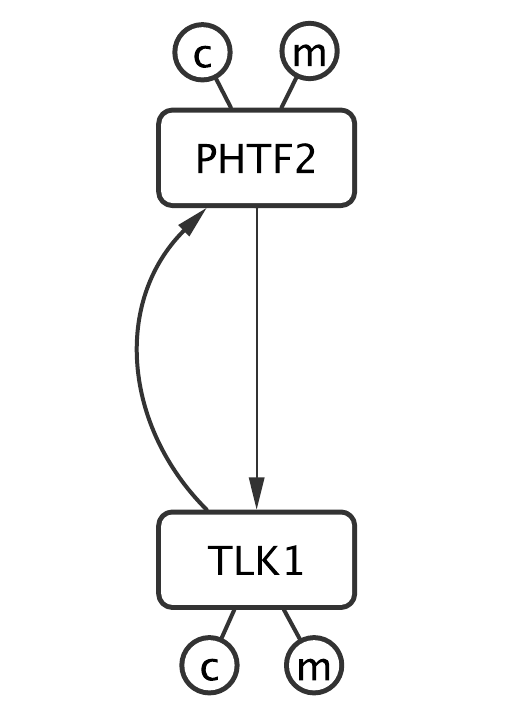}\label{zod3}}
	\subfloat[COAD]{\includegraphics[width=0.2\textwidth]{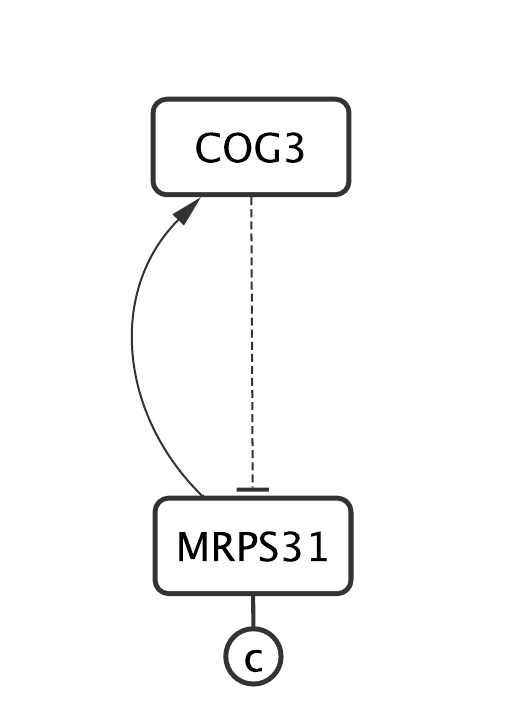}\label{zod1}} 
	\caption{Interactions of three pairs of genes in three cancers. Genes attached by tags ``c" and ``m" indicate that they are associated with copy number and methylation, respectively. Solid lines with arrowheads represent stimulatory interactions, whereas dashed lines with horizontal bars denote inhibitory regulation. Edge width is proportional to its posterior probability. }\label{fig:zod}
\end{figure}

% We leave the full network analysis to Section \ref{sec:capa}.\\ 
Among the 1395 gene pairs, 518 pairs
are estimated to have one-way
interactions, 131 pairs have two-way interactions and the rest have no
interactions. In addition, we find 2544 intragenic GE-CN interactions
and 1157 intragenic GE-ME interactions. We show three gene pairs
in Figure \ref{fig:zod}. These interactions are confirmed by ZODIAC
\citep{zhu2015zodiac} but our approach has richer interpretations. In
Figure \ref{zod2}, both our method and ZODIAC find a negative
interaction between MEA1 and UBR2 whereas our analysis indicates
that  MEA1 downregulates UBR2 specifically in rectum adenocarcinoma
(READ). Similarly, in Figure \ref{zod3}, while ZODIAC finds positive
interaction between PHTF2 and TLK1, our model suggests a positive
feedback loop between these two genes in testicular germ cell tumors
(TGCT). Figure \ref{zod1} manifests a unique feature of our model, a
negative feedback loop between COG3 and MRPS31 in colon adenocarcinoma
(COAD). Negative feedback loops are a common and important mechanism in
gene regulations. For example, it can reduce expression 
noise and intercellular variability in molecular levels
\citep{singh2011negative}. But negative feedback loops are not allowed by
DAGs and MRFs, both of which include the simplified assumption
that gene interaction is always symmetric, either positive or
negative. Our model is able to generate hypotheses that involve both
one-way and two-way interactions and both positive and negative
feedback loops.  
\subsection{Colon adenocarcinoma pathway analysis}
\label{sec:capa}
\begin{figure}[h]
	\def\tabularxcolumn#1{m{#1}}
	\begin{tabularx}{\linewidth}{@{}cXX@{}}
		\multirow{-9}[2]{*}{\subfloat[Full network]{\includegraphics[width=.45\textwidth,height=.45\textwidth]{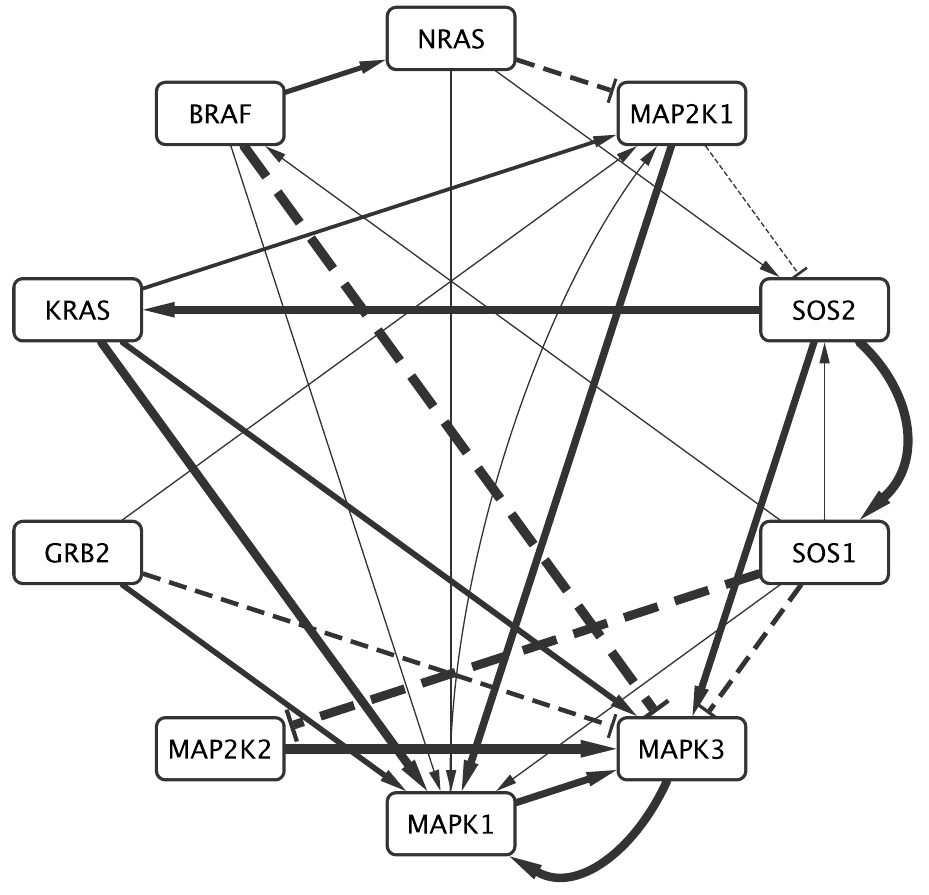}\label{fig:fn}}}
		\begin{tabular}{cc}
			\subfloat[Feed-forward loop]{\includegraphics[width=.25\textwidth]{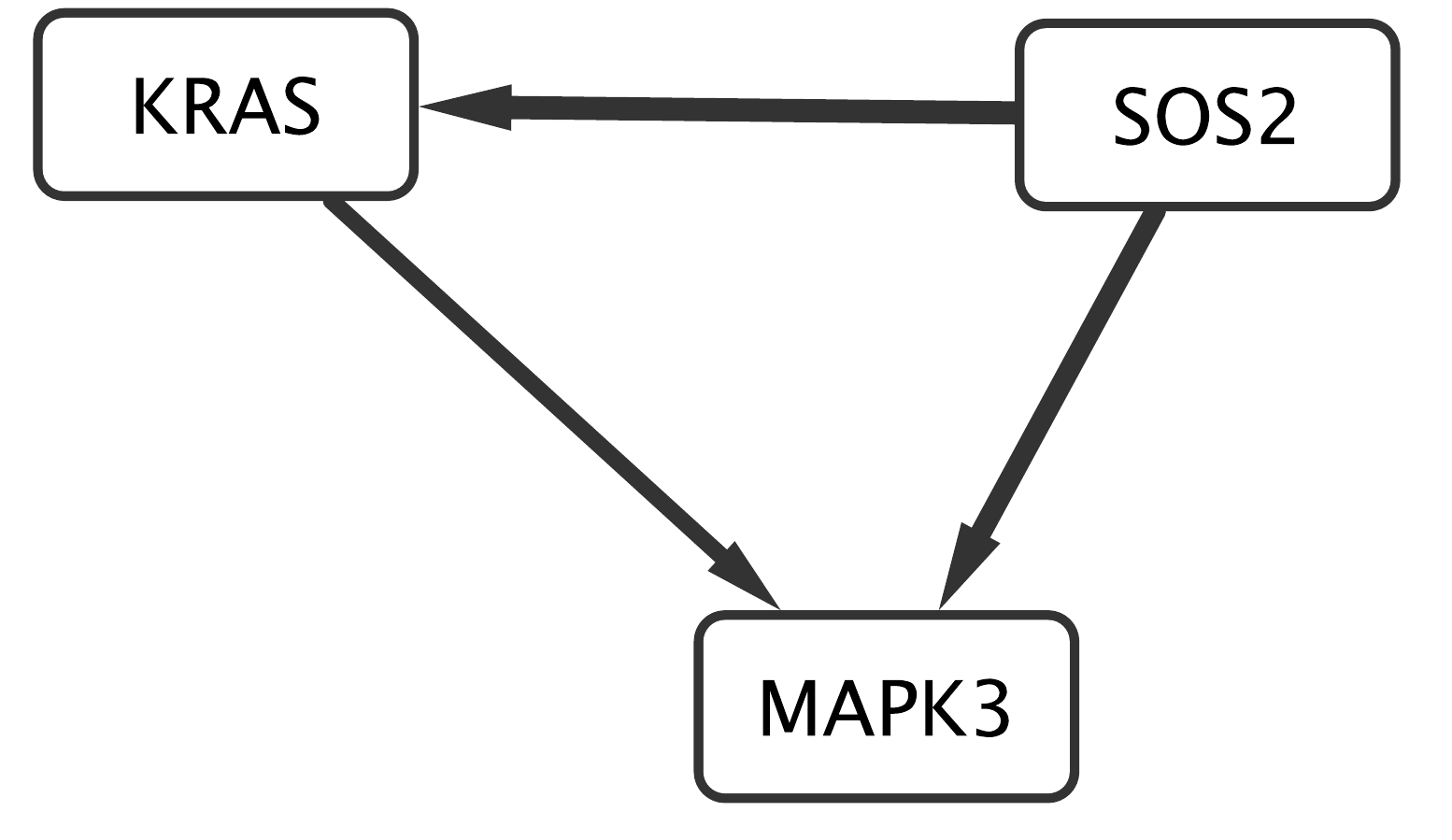}\label{cfl}} 
			& \subfloat[Negative feedback loop]{\includegraphics[width=.25\textwidth]{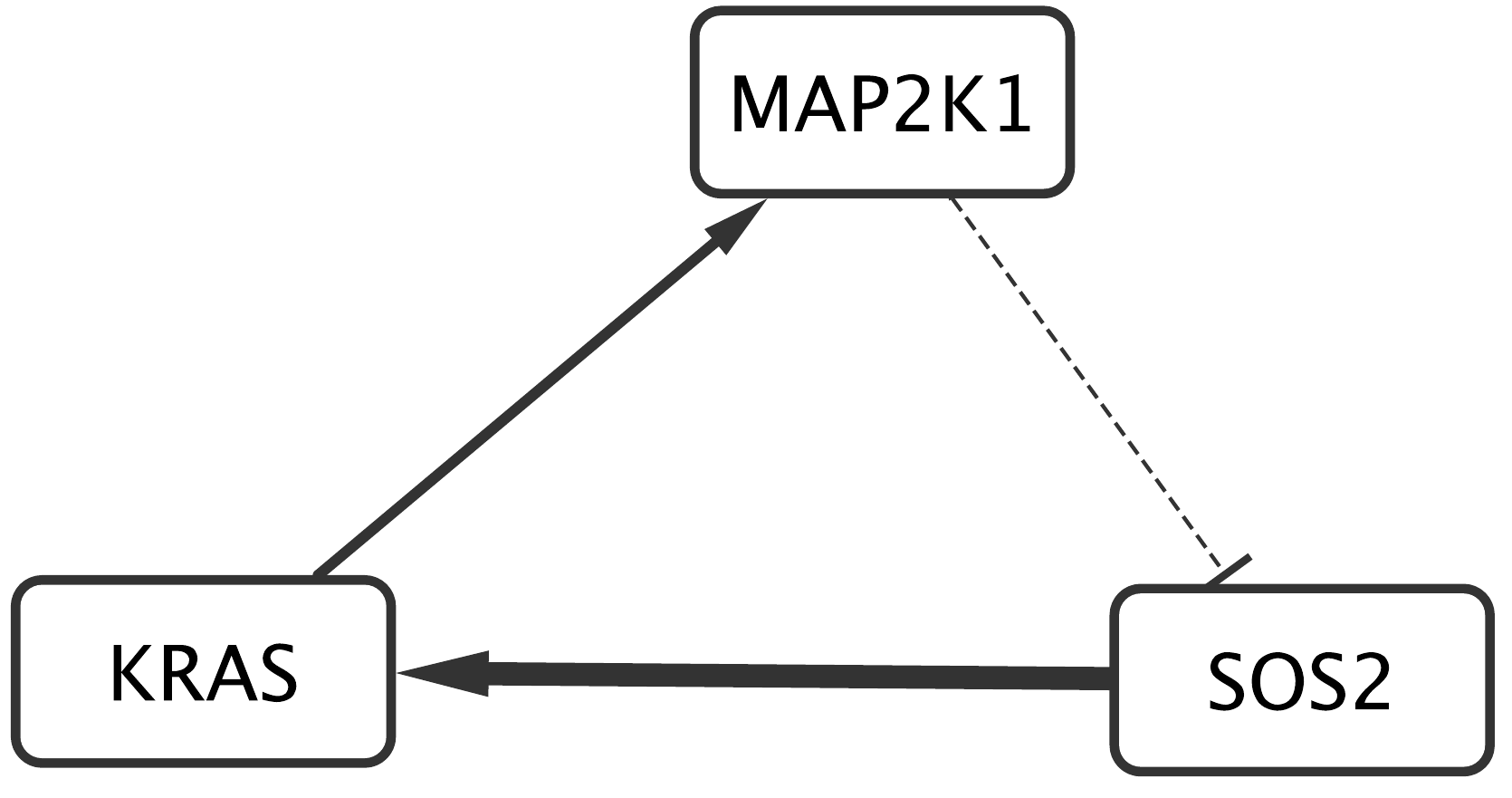}\label{nfbl}}\\
			\subfloat[Positive feedback loop]{\includegraphics[width=.25\textwidth]{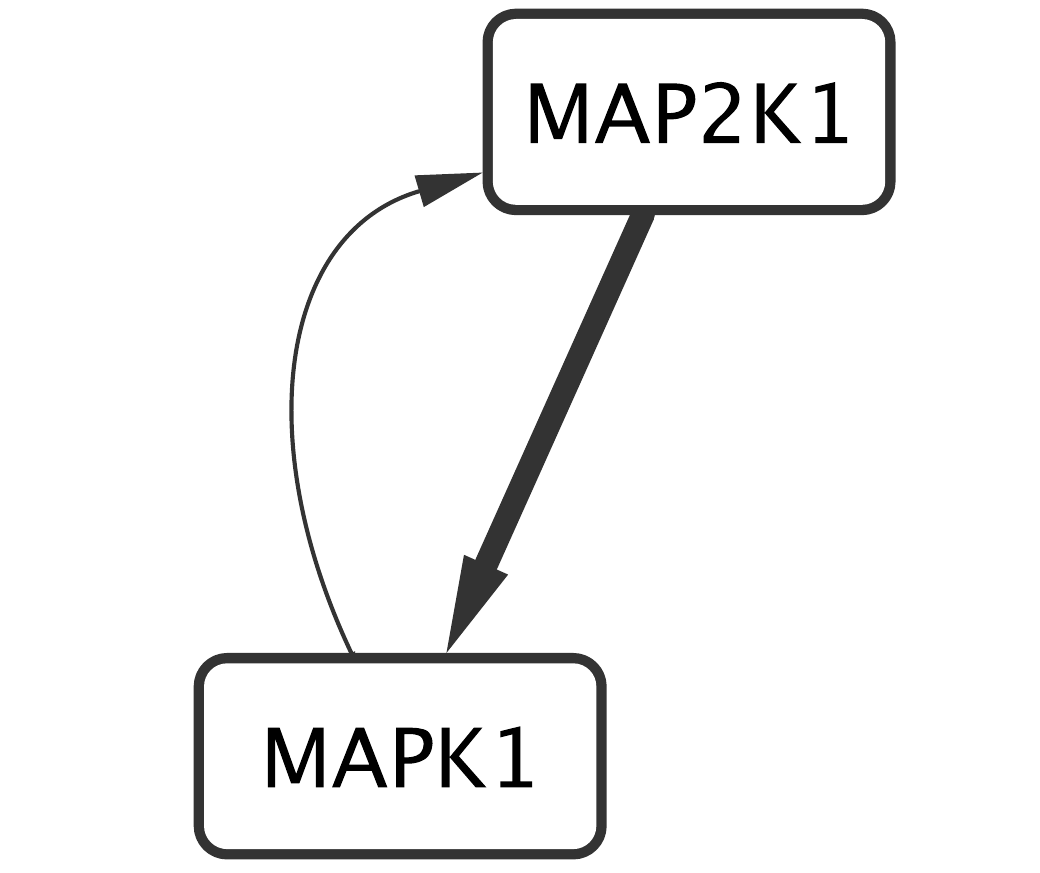}\label{pfbl}}
			& \subfloat[Cascade]{\includegraphics[width=.25\textwidth]{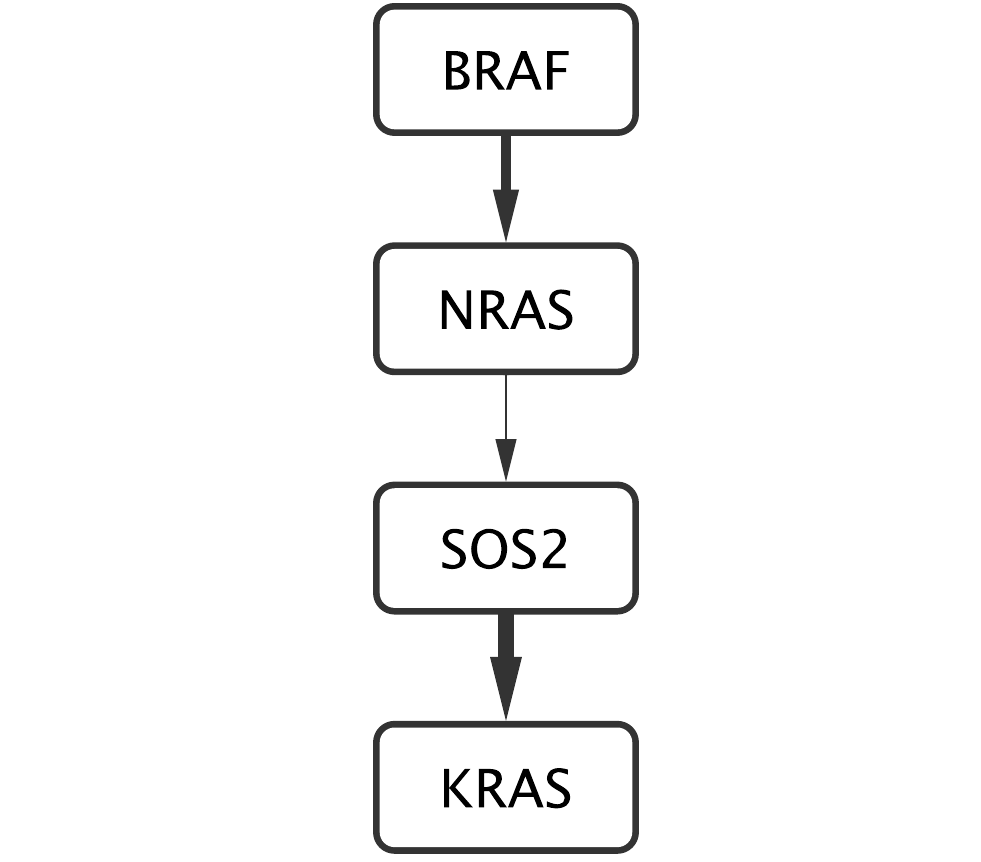}\label{cas}}\\
		\end{tabular}
	\end{tabularx}
	\caption{Estimated gene network for colon adenocarcinoma. Solid lines with arrowheads represent stimulatory interactions, whereas dashed lines with horizontal bars denote inhibitory regulation. Line width is proportional to its posterior probability. Panel (a) full network; Panels (b)-(e) network motifs.}\label{fig:grn}
\end{figure}

We analyze data from colon adenocarcinoma  (COAD) samples. We focus on genes that
are mapped to the RAS-MAPK pathway, which is critical for initiation of
carcinogenesis in COAD \citep{colussi2013molecular}. The RAS-MAPK pathway
includes $p=10$ core genes.
Restricting to samples with available 
GE, CN and ME data, 
the sample size is $n=276$. We run two parallel MCMC simulations, each with
50,000 iterations, discard the first 50\% as burn-in and thin the
chains to every 5th sample. 
MCMC diagnostics show no evidence for lack of practical convergence 
(for details see part C of the Supplementary
Materials). We summarize the posterior distribution on the unknown
graph by controlling the posterior expected
$\mbox{FDR}<10\%$.\\
\indent We find that all genes are associated with their respective copy
number and NRAS, MAP2K1, MAPK1 and MAPK3 are associated with
methylation as well. Therefore, the gene interactions are fully
identifiable by the argument in the last paragraph of Section
\ref{sec:rgm}. The full network of gene interactions is shown in
Figure \ref{fig:fn}. The solid lines with arrowheads represent
stimulatory interactions and the dashed lines with perpendicular bars
denote inhibitory regulations. Uncertainties about the selected edges
are quantified as posterior inclusion probabilities which are 
displayed as edge width in the figure. In total, we find 21
stimulatory and 7 inhibitory regulatory relationships.

\indent Gene networks are often made up of a small set of recurrent
regulation patterns, called network motifs, which can be thought of as
fundamental building blocks for the network and are expected to occur
more often in gene networks than in random networks. In Figures
(\ref{cfl})-(\ref{cas}), we display four motifs identified by our
method that are commonly observed in gene networks
\citep{alon2007network}. Figure (\ref{cfl}) shows a feed-forward loop
among SOS2, KRAS and MAPK3. Part of this feed-forward loop is well
studied in COAD \citep{zenonos2013ras}. SOS2 binds KRAS and removes
guanosine diphosphate (GDP) molecules from KRAS and thus allows
guanosine triphosphate (GTP) molecules to bind and activate it. The
active KRAS would eventually activates MAPK3 through the kinase
cascade. Another important motif is feedback loop. We present a
negative and a positive feedback loop in Figures \ref{nfbl} and
\ref{pfbl}, respectively. The regulatory relationships in the negative
feedback loop of SOS2, KRAS and MAP2K1 have been studied
extensively. SOS2$\rightarrow$KRAS$\rightarrow$MAP2K1 is part of the
well-known MAP kinases cascade
\citep{plotnikov2011mapk,zenonos2013ras} while MAP2K1 can
phosphorylate and inhibit SOS2 and thereby reduces MAP2K1 activation
\citep{holt1996insulin,mendoza2011ras}. For the positive feedback loop
(Figure \ref{pfbl}), MAP2K1 activating MAPK1 is again part of the MAP
kinases cascade but the reversed activation is less explored in the
literature. Similarly, another network motif (Figure \ref{cas}),
regulatory cascade, also need to be validated by further biological
experiment.

\begin{figure}[h]
	\centering
	\subfloat[]{\includegraphics[width=.7\textwidth]{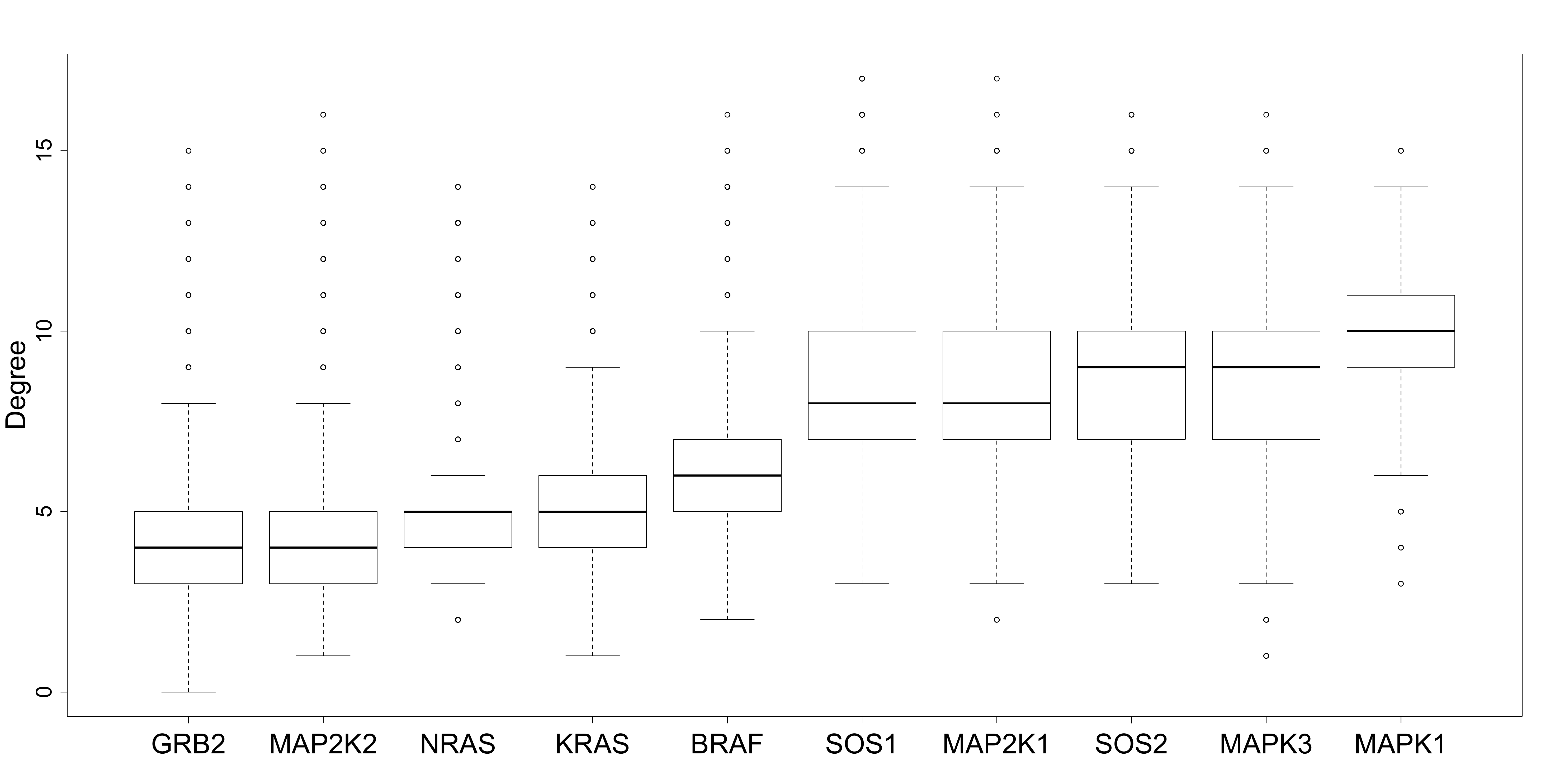}\label{fig:deg}}
	\subfloat[]{\includegraphics[width=.21\textwidth]{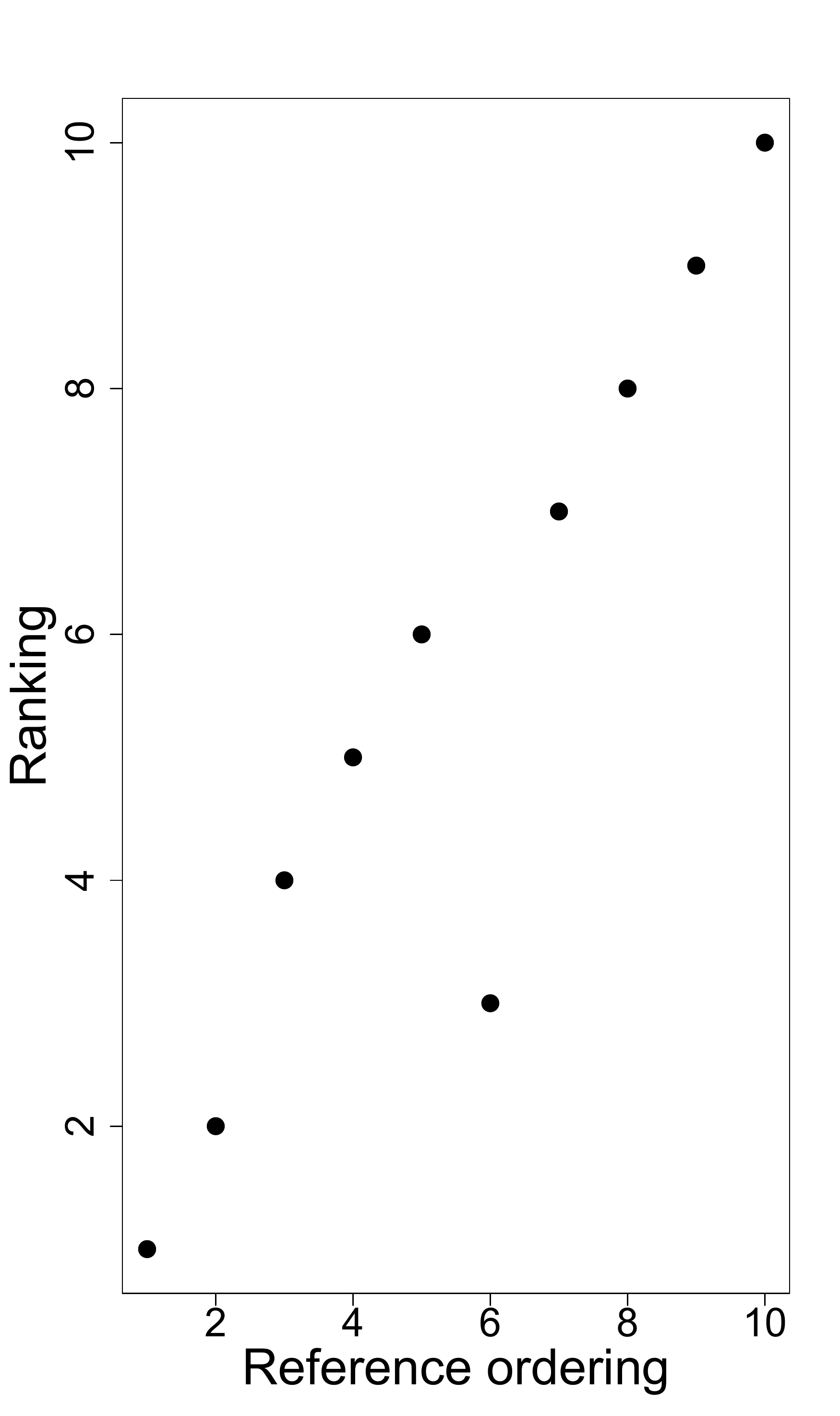}\label{fig:ranking}}
	\caption{Panel (a): box plots for posterior distribution of
          degree of each gene; Panel (b): scatter plot for ranking vs
          reference ordering. Ranking of the genes is based on the
        Score$_j$ defined by \eqref{score}. The reference ordering is
        based on \eqref{casc} where the genes with the same bracket are
        assigned arbitrary ordering. }
\end{figure}

On the gene level, we calculate the posterior distribution of
degree of each gene, that is, the number of edges connected to the
gene. We visualize these posterior distributions by box plots in
Figure \ref{fig:deg}. Highly connected genes are often called hub
genes which are usually more involved in multiple regulatory
activities than non-hub genes. MAPK1 and MAPK3 appear to be the two
most highly connected genes. Several studies have shown that
overexpression of MAPK1/3 plays an critical role in the progression of
COAD \citep{fang2005mapk} and is responsible for proliferation,
differentiation, survival, migration and angiogenesis of tumors in
many cancers \citep{dhillon2007map}. \\

For another quantitative exploration of the known MAPK signalling
cascade we consider the following inference.
The following signaling pathway/cascade has been extensively studied
and validated in biological literatue
\citep{plotnikov2011mapk,zenonos2013ras}, 
\begin{eqnarray}
  \label{casc}
  \left\{\begin{array}{c}\mbox{GRB2}\\\mbox{SOS1}\\\mbox{SOS2}\end{array}\right\}\longrightarrow\left\{\begin{array}{c}\mbox{NRAS}\\\mbox{KRAS}\end{array}\right\}\longrightarrow \mbox{BRAF}\longrightarrow\left\{\begin{array}{c}\mbox{MAP2K1}\\\mbox{MAP2K2}\end{array}\right\}\longrightarrow\left\{\begin{array}{c}\mbox{MAPK1}\\\mbox{MAPK3}\end{array}\right\}
\end{eqnarray}
where genes within each curly brace belong to the same gene family
(except for GRB2 for which the protein often binds to SOS to form a
protein complex) and play similar roles in the pathway. We label each
gene in cascade (\ref{casc}) from left to right with integer
$j=1,\dots,p$ as our reference ordering. The labels within each curly
brace are arbitrary. To compare our findings with this well
established cascade, we score each gene $j$ by  
\begin{equation}
\mbox{Score}_j=\mbox{indegree}_j-\mbox{outdegree}_j, \label{score}
\end{equation}
where $\mbox{indegree}_j$ is the number of edges pointing towards gene
$j$ and $\mbox{outdegree}_j$ is the number of edges pointing away from
gene $j$. Intuitively, gene $j$ is likely to be on the left of gene
$k$ in cascade (\ref{casc}) if $\mbox{Score}_j<\mbox{Score}_k$.  We
then rank the scores in an increasing order and plot the rank against
the reference ordering in Figure \ref{fig:ranking}. We also calculate
the normalized Kendall's tau distance, which is the ratio of the
number of discordant pairs over the total number of pairs. The
normalized Kendall's tau distance lies between 0 and 1, with 0
indicating a perfect agreement of the two orderings and 1 indicating a
perfect disagreement. The resulting Kendall's tau distance is
0.07. Our findings appear to be consistent with the biologically
validated pathway since both Figure \ref{fig:ranking} and the
Kendall's tau distance indicate a good concordance between our
rankings and the reference ordering. 
%\note{I think the results are amazingly good! MAPK signaling pathway
%  can be summarized as GRB2 --> SOS --> RAS --> MEK (MAP2K) -->
%  MAPK. If we see the directions of Figure 5(a), other than that we
%  miss GRB2 --> SOS, we almost got everything else right! I might even
%  think about giving a scoring system: for each gene, if it is on the
%  receiving end of an edge, it gets +1 point; if it's on the departing
%  end, it gets -1 point; otherwise, no point. So MAPK1 and MAPK3 both
%  have the largest points, indicating they are at the end of the
%  signaling cascade. In contrast, SOS, RAS, genes have mostly negative
%  points indicating they are at the beginning of the pathway. We could
%  add these in needed. I just think the results make a lot of sense!! --Yuan}
%\begin{figure}[h]
%	\centering
%	\includegraphics[width=.5\textwidth]{ras_fdr10.pdf}
%	\caption{Estimated gene network for colon adenocarcinoma. Solid lines with arrowheads represent stimulatory interactions, whereas dashed lines with horizontal bars denote inhibitory regulation. Line width is proportional to its posterior probability.}\label{fig:grn}
%\end{figure}
%\begin{figure}
%	\centering
%	\subfigure[]{\includegraphics[width=0.2\textwidth]{g3.pdf}\label{tg1}} 
%	\subfigure[]{\includegraphics[width=0.2\textwidth]{g1.pdf}}
%	\subfigure[]{\includegraphics[width=0.2\textwidth]{g2.pdf}} 
%	\subfigure[]{\includegraphics[width=0.2\textwidth]{g4.pdf}}
%	\caption{}\label{twogenes}
%\end{figure}
\section{Discussion}
\label{sec:disc}
In this article, we have introduced a Gaussian RGM to model gene
regulatory relationships from genomic data. RGMs are statistically
more general and biologically more interpretable than MRFs and
DAGs. By integrating DNA level information, we are able to
differentiate between RGMs that belong to the same Markov equivalence
class. We exploit the connection between RGMs and SEMs for efficient
inference. We constructed a prior probability model for the
unknown graph using a thresholded model which marginally defines a mixture of non-local prior and a
point mass. We use simulation studies to illustrate the performance of
our method in terms of graph structure learning. Our method is applied
to a ZODIAC gene interaction analysis and a colon cancer pathway
analysis. Some of our findings are consistent with the literature,
while others need to be validated by biological experiments. Although
our applications focus on GRNs, the proposed approach is general and
can be potentially applied to other scientific settings such as
climate sciences and macroeconomics. The approach works for any network where some edges have known
direction if included (inclusion itself need not be fixed). 

The link between RGMs and SEMs is based on the assumption that the
gene expressions are multivariate Gaussian, which could be thought of
as one limitation of our model. We have empirically
demonstrated the robustness of the proposed model against slight model
misspecification. More general sampling models might need an additional
hierarchical layer of latent variables, such as latent probit scores
for binary outcomes. 
% , the theoretical work that (dis)proves equivalence
% between RGMs and SEMs with non-Gaussian data is still needed to be
% developed. We would like to pursue this direction in our future work.  

%\section*{Acknowledgement}

%Yang Ni, Peter M\"uller and Yuan Ji's research were partially supported by NIH/NCI grant a RO1 CA 083932.

\bibliography{rgm_ref}

\begin{thebibliography}{}

\bibitem[Alon, 2007]{alon2007network}
Alon, U. (2007).
\newblock Network motifs: theory and experimental approaches.
\newblock {\em Nature Reviews Genetics}, 8(6):450--461.

\bibitem[Colussi et~al., 2013]{colussi2013molecular}
Colussi, D., Brandi, G., Bazzoli, F., and Ricciardiello, L. (2013).
\newblock Molecular pathways involved in colorectal cancer: implications for
  disease behavior and prevention.
\newblock {\em International journal of molecular sciences},
  14(8):16365--16385.

\bibitem[Dhillon et~al., 2007]{dhillon2007map}
Dhillon, A.~S., Hagan, S., Rath, O., and Kolch, W. (2007).
\newblock Map kinase signalling pathways in cancer.
\newblock {\em Oncogene}, 26(22):3279--3290.

\bibitem[Dobra et~al., 2012]{dobra2012bayesian}
Dobra, A., Lenkoski, A., and Rodriguez, A. (2012).
\newblock Bayesian inference for general gaussian graphical models with
  application to multivariate lattice data.
\newblock {\em Journal of the American Statistical Association}.

\bibitem[Fang and Richardson, 2005]{fang2005mapk}
Fang, J.~Y. and Richardson, B.~C. (2005).
\newblock The mapk signalling pathways and colorectal cancer.
\newblock {\em The lancet oncology}, 6(5):322--327.

\bibitem[Goldenberg et~al., 2010]{goldenberg2010survey}
Goldenberg, A., Zheng, A.~X., Fienberg, S.~E., and Airoldi, E.~M. (2010).
\newblock A survey of statistical network models.
\newblock {\em Foundations and Trends{\textregistered} in Machine Learning},
  2(2):129--233.

\bibitem[Green and Thomas, 2013]{green2013sampling}
Green, P.~J. and Thomas, A. (2013).
\newblock Sampling decomposable graphs using a markov chain on junction trees.
\newblock {\em Biometrika}, 100(1):91--110.

\bibitem[Hoff, 2009]{hoff_2009_cmot}
Hoff, P. (2009).
\newblock Multiplicative latent factor models for description and prediction of
  social networks.
\newblock {\em Computational \& Mathematical Organization Theory},
  15(4):261--272.

\bibitem[Holt et~al., 1996]{holt1996insulin}
Holt, K.~H., Kasson, B.~G., and Pessin, J.~E. (1996).
\newblock Insulin stimulation of a mek-dependent but erk-independent sos
  protein kinase.
\newblock {\em Molecular and Cellular Biology}, 16(2):577--583.

\bibitem[Johnson and Rossell, 2010]{johnson2010use}
Johnson, V.~E. and Rossell, D. (2010).
\newblock On the use of non-local prior densities in bayesian hypothesis tests.
\newblock {\em Journal of the Royal Statistical Society: Series B (Statistical
  Methodology)}, 72(2):143--170.

\bibitem[Koster, 1996]{koster1996markov}
Koster, J.~T. (1996).
\newblock Markov properties of nonrecursive causal models.
\newblock {\em The Annals of Statistics}, 24(5):2148--2177.

\bibitem[Mendoza et~al., 2011]{mendoza2011ras}
Mendoza, M.~C., Er, E.~E., and Blenis, J. (2011).
\newblock The ras-erk and pi3k-mtor pathways: cross-talk and compensation.
\newblock {\em Trends in biochemical sciences}, 36(6):320--328.

\bibitem[Mitra et~al., 2013]{mitra2013bayesian}
Mitra, R., M{\"u}ller, P., Liang, S., Yue, L., and Ji, Y. (2013).
\newblock A bayesian graphical model for chip-seq data on histone
  modifications.
\newblock {\em Journal of the American Statistical Association},
  108(501):69--80.

\bibitem[M\"uller et~al., 2006]{muller2006fdr}
M\"uller, P., Parmigiani, G., and Rice, K. (2006).
\newblock Fdr and bayesian multiple comparisons rules.

\bibitem[Newton et~al., 2004]{newton2004detecting}
Newton, M.~A., Noueiry, A., Sarkar, D., and Ahlquist, P. (2004).
\newblock Detecting differential gene expression with a semiparametric
  hierarchical mixture method.
\newblock {\em Biostatistics}, 5(2):155--176.

\bibitem[Ni et~al., 2016]{ni2016}
Ni, Y., Stingo, F., and Baladandayuthapani, V. (2016).
\newblock Bayesian graphical regression.
\newblock {\em submitted}.

\bibitem[Ni et~al., 2015]{ni2015bayesian}
Ni, Y., Stingo, F.~C., and Baladandayuthapani, V. (2015).
\newblock Bayesian nonlinear model selection for gene regulatory networks.
\newblock {\em Biometrics}, 71(3):585--595.

\bibitem[Plotnikov et~al., 2011]{plotnikov2011mapk}
Plotnikov, A., Zehorai, E., Procaccia, S., and Seger, R. (2011).
\newblock The mapk cascades: signaling components, nuclear roles and mechanisms
  of nuclear translocation.
\newblock {\em Biochimica et Biophysica Acta (BBA)-Molecular Cell Research},
  1813(9):1619--1633.

\bibitem[Shin et~al., 2010]{shin2010functional}
Shin, S.-Y., Rath, O., Zebisch, A., Choo, S.-M., Kolch, W., and Cho, K.-H.
  (2010).
\newblock Functional roles of multiple feedback loops in erk and wnt signaling
  pathways that regulate epithelial-mesenchymal transition.
\newblock {\em Cancer research}, 70(17):6715.

\bibitem[Singh, 2011]{singh2011negative}
Singh, A. (2011).
\newblock Negative feedback through mrna provides the best control of
  gene-expression noise.
\newblock {\em NanoBioscience, IEEE Transactions on}, 10(3):194--200.

\bibitem[Spirtes, 1995]{spirtes1995directed}
Spirtes, P. (1995).
\newblock Directed cyclic graphical representations of feedback models.
\newblock In {\em Proceedings of the Eleventh conference on Uncertainty in
  artificial intelligence}, pages 491--498. Morgan Kaufmann Publishers Inc.

\bibitem[Stingo et~al., 2010]{stingo2010bayesian}
Stingo, F.~C., Chen, Y.~A., Vannucci, M., Barrier, M., and Mirkes, P.~E.
  (2010).
\newblock A bayesian graphical modeling approach to microrna regulatory network
  inference.
\newblock {\em The annals of applied statistics}, 4(4):2024.

\bibitem[TCGA, 2012]{cancer2012comprehensive}
TCGA (2012).
\newblock Comprehensive molecular characterization of human colon and rectal
  cancer.
\newblock {\em Nature}, 487(7407):330--337.

\bibitem[Telesca et~al., 2012a]{telesca2012modeling2}
Telesca, D., M{\"u}ller, P., Kornblau, S.~M., Suchard, M.~A., and Ji, Y.
  (2012a).
\newblock Modeling protein expression and protein signaling pathways.
\newblock {\em Journal of the American Statistical Association},
  107(500):1372--1384.

\bibitem[Telesca et~al., 2012b]{telesca2012modeling1}
Telesca, D., M{\"u}ller, P., Parmigiani, G., and Freedman, R.~S. (2012b).
\newblock Modeling dependent gene expression.
\newblock {\em The Annals of Applied Statistics}, 6(2):542--560.

\bibitem[Wang and West, 2009]{wang2009bayesian}
Wang, H. and West, M. (2009).
\newblock Bayesian analysis of matrix normal graphical models.
\newblock {\em Biometrika}, 96(4):821--834.

\bibitem[Wang et~al., 2013]{Wang2013}
Wang, W., Baladandayuthapani, V., Holmes, C.~C., and Do, K.-A. (2013).
\newblock Integrative network-based bayesian analysis of diverse genomics data.
\newblock {\em BMC Bioinformatics}, 14(Suppl 13):S8.

\bibitem[Yajima et~al., 2015]{yajima2015detecting}
Yajima, M., Telesca, D., Ji, Y., and M{\"u}ller, P. (2015).
\newblock Detecting differential patterns of interaction in molecular pathways.
\newblock {\em Biostatistics}, 16(2):240--251.

\bibitem[Zenonos and Kyprianou, 2013]{zenonos2013ras}
Zenonos, K. and Kyprianou, K. (2013).
\newblock Ras signaling pathways, mutations and their role in colorectal
  cancer.
\newblock {\em World J Gastrointest Oncol}, 5(5):97--101.

\bibitem[Zhang et~al., 2005]{zhang2005hierarchical}
Zhang, D., Wells, M.~T., Turnbull, B.~W., Sparrow, D., and Cassano, P.~A.
  (2005).
\newblock Hierarchical graphical models: An application to pulmonary function
  and cholesterol levels in the normative aging study.
\newblock {\em Journal of the American Statistical Association},
  100(471):719--727.

\bibitem[Zhu et~al., 2014]{zhu2014tcga}
Zhu, Y., Qiu, P., and Ji, Y. (2014).
\newblock Tcga-assembler: open-source software for retrieving and processing
  tcga data.
\newblock {\em Nature methods}, 11(6):599--600.

\bibitem[Zhu et~al., 2015]{zhu2015zodiac}
Zhu, Y., Xu, Y., Helseth, D.~L., Gulukota, K., Yang, S., Pesce, L.~L., Mitra,
  R., M{\"u}ller, P., Sengupta, S., Guo, W., et~al. (2015).
\newblock Zodiac: a comprehensive depiction of genetic interactions in cancer
  by integrating tcga data.
\newblock {\em Journal of the National Cancer Institute}, 107(8):djv129.

\end{thebibliography}

\end{document}